\documentclass[12pt]{article}
\usepackage{amssymb}

\usepackage[dvips]{graphicx, color}

\newcommand{\ncm}{\newcommand}
\newcommand{\rencm}{\renewcommand}
 
\def\a{\alpha}

\def\f{\varphi}    %\phi
\def\l{\lambda}
\def\m{\mu}

\def\o{\omega}
\def\p{\pi}       % Also, \varpi
\def\r{\rho}      %     \varrho
\def\s{\sigma}    %     \varsigma
\def\t{\tau}

\def\D{\Delta}

\def\ph{\hat{\p}}
\def\fh{\hat{\f}}

\ncm{\dsp}{\displaystyle}
\ncm{\nn}{\nonumber}
\ncm{\nnn}{\nonumber\linebreak[4]}
\ncm{\nit}{\noindent}
\ncm{\del}{\partial}
\ncm{\av}[1]{\mbox{$\langle #1 \rangle$}}
\ncm{\ket}[1]{\mbox{$| #1 \rangle$}}
\ncm{\bra}[1]{\mbox{$\langle #1 |$}}
\ncm{\avc}[1]{\mbox{$\langle #1 \rangle_{\psi}$}}
\ncm{\half}{\mbox{{\small $\frac{1}{2}$}} }
\ncm{\quart}{\mbox{{\small $\frac{1}{4}$}} }
\ncm{\tq}{\mbox{{\small $\frac{3}{4}$}} }
\ncm{\third}{\mbox{{\small $\frac{1}{3}$}} }
\ncm{\sixth}{\mbox{{\small $\frac{1}{6}$}} }
\ncm{\eigth}{\mbox{{\small $\frac{1}{8}$}} }
\ncm{\thrhalf}{\mbox{{\small $\frac{3}{2}$}} }
\ncm{\thrfor}{\mbox{{\small $\frac{3}{4}$}} }
\ncm{\twothi}{\mbox{{\small $\frac{2}{3}$}} }
\ncm{\fivtwo}{\mbox{{\small $\frac{5}{2}$}} }

\ncm{\dxx}{\mbox{$\partial_{x}^2$}}
\ncm{\dx}{\mbox{$\partial_{x}$}}
\ncm{\dt}{\mbox{$\partial_{t}$}}
\ncm{\dtt}{\mbox{$\partial_{t}^2$}}
\ncm{\un}{1\!\!1}
\ncm{\RE}{\mbox{Re}}
\ncm{\IM}{\mbox{Im}}
\ncm{\Tr}{\mbox{tr}\,}
\ncm{\diag}{\mbox{diag}\,}
\ncm{\Det}{\mbox{Det}\,}
\ncm{\Log}{\mbox{Log}\,}
\ncm{\ra}{\rightarrow}
\ncm{\la}{\leftarrow}
\ncm{\dg}{\dagger}
\ncm{\pr}{\prime}
\ncm{\ha}{\hat{a}}
\ncm{\hP}{\hat{P}}

\ncm{\aplt}{ \mbox{}_{\textstyle \sim}^{\textstyle < }     }
\ncm{\apgt}{ \mbox{}_{\textstyle \sim}^{\textstyle > }     }
\ncm{\Oa}{\mbox{$\mbox{O}(a)$}}
\ncm{\Sp}{\mbox{\hspace{1.0cm}}}
\ncm{\capit}[1]{\caption{\it #1}}

\def\be{\begin{equation}}
\def\ee{\end{equation}}
\def\bea{\begin{eqnarray}}
\def\eea{\end{eqnarray}}
\def\bi{\begin{itemize} \itemsep = 0.01\itemsep  }
\def\bii{\begin{itemize} \small \itemsep = 0.01\itemsep }
\def\ei{\end{itemize}}
\def\bc{\begin{center}}
\def\ec{\end{center}}
\def\bs{\begin{slide}}
\def\es{\end{slide}}

\def\beac{\begin{eqnarray} \color [rgb] {0,0,1} }
\def\eeac{\end{eqnarray} }

\ncm{\shead}[1]{\bc { \Large \color [rgb]{1.0, .0, .0} #1 \normalcolor} \ec}
\ncm{\ssubh}[1]{{\large \color [rgb]{1.0,.0,.1} #1 \normalcolor}}
\rencm{\thefootnote}{\mbox{\protect{$\fnsymbol{footnote}$}} }

\ncm{\front}[5]   %% date, ITFA#, title, authors, abstract
{
   \begin{titlepage}
      \noindent {#1} \hfill {#2}\\
      \begin{center}
         \vspace{1.5\baselineskip}
         {\Large\bf  #3  } \\
         \vspace{2\baselineskip}
         \vspace{1.5\baselineskip}
          #4\\
         \vspace{1.5\baselineskip}
   
         Institute of Theoretical Physics, University of Amsterdam, \\
         Valckenierstraat 65, 1018 XE Amsterdam,
         The~Netherlands.
    
      \end{center}
      \vfill
      {\bf Abstract}\\
       #5
   \end{titlepage} 
}

\ncm{\frontslide}[4]
{
   \begin{titlepage}
      \noindent {#1} \hfill {#2}\\
      \begin{center}
         \vspace{1.5\baselineskip}
         {\Large\bf  #3  } \\
         \vspace{2\baselineskip}
         \vspace{1.5\baselineskip}
          #4\\
         \vspace{1.5\baselineskip}
   
         Institute of Theoretical Physics, \\
         Valckenierstraat 65, 1018 XE Amsterdam,
         The~Netherlands.
    
      \end{center}
   \end{titlepage} 
}

\newcommand{\vecx}{{\bf x}}
\newcommand{\veck}{{\bf k}}

\begin{document}

%title page:  date, ITFA#, title, authors, abstract

\front
{
   August 2001
}
{
   ITFA-2000-22
}
{
  {Staying Thermal with Hartree Ensemble Approximations }
}
{  
   Mischa Sall\'e\footnote{e-mail: msalle@science.uva.nl}
   Jan Smit\footnote{e-mail: jsmit@science.uva.nl}
   and Jeroen C. Vink\footnote{e-mail: jcvink@science.uva.nl}
}
{
We study thermal behavior of a recently introduced 
Hartree ensemble approximation, which allows for 
non-perturbative inhomogeneous field configurations 
as well as for approximate thermalization, in the 
$\varphi^4$ model in 1+1 dimensions. 
Using ensembles with a free field thermal distribution 
as out-of-equilibrium initial conditions we determine 
thermalization time scales. The time scale for which
the system stays in approximate quantum thermal 
equilibrium is an indication of the time scales for 
which the approximation method stays reasonable.
This time scale turns out to be two orders of magnitude 
larger than the time scale for thermalization,
in the range of couplings and temperatures studied. 
We also discuss simplifications of our method which 
are numerically more efficient and make a comparison 
with classical dynamics.
}

\section{Introduction}
\label{introduction}

It is highly desirable to be able to numerically simulate quantum
field dynamics in real time. 
This will give an important tool for the study of non-perturbative phenomena in 
out-of-equilibrium systems, such as phase transitions
in the early universe or the quark-hadron transition
in heavy-ion collisions.
Using real time dynamics may also offer an alternative
to simulating equilibrium physics, just like molecular
dynamics simulations provide a fruitful alternative to Monte Carlo 
simulations in other areas of physics. 
Simulating quantum fields in real time is very difficult. Direct approaches, 
such as solving the Schr\"odinger equation for the field wave-functional 
or evaluating the Minkowski path integral using Monte Carlo methods, 
are prohibitively time-consuming. 
One has to resort to approximate methods, of which the classical
approximation (
%see e.g.\ [1--6]),
see e.g.\ [1--10]),
\nocite{MoRu00,SmTa99,afterpreh_1,afterpreh_2,afterpreh_3,newewbary_1,newewbary_2,AaBo00,AaSm98_1,AaSm98_2}
the Hartree approximation
and large $n$ methods (see e.g.~\cite{CoHa94,BoVe00,Miea00}) 
are most commonly used. 

In the classical approximation one assumes that the fields follow
classical equations of motion. This
is reasonable when the occupation numbers of the field quanta 
are large, but in field theory this is never the case for all modes. 
For instance, at high temperatures 
the low momentum modes of the fields are highly occupied and follow
the classical Boltzmann distribution,
but at large momenta occupation numbers are low and the classical distribution
differs significantly from the quantum Bose-Einstein distribution, 
giving rise to Rayleigh-Jeans divergences.

These divergences are absent in the  
Hartree and large $n$ approximations, which include quantum effects in the
field dynamics.
In these approximations the density operator is effectively gaussian,
such that
all information is contained in the mean field and two-point function.
These are usually assumed to be translationally invariant.
A problem which arises under these circumstances is that the system
does not thermalize \cite{CoHa94,BoVe00}, 
in contrast to the classical
approximation which has no such problem \cite{AaBo00}. 
The particles corresponding to the 
quantum modes of the field
interact via the mean field and since this field is homogeneous 
there is no scattering which leads to redistribution of occupation
numbers over different momentum modes
and there is no thermalization. 
One way to amend this situation may be an improved approximation 
which includes direct scattering \cite{BeCo00}.

Recently we have extended the Hartree approximation by writing an initial
density operator as an ensemble of coherent states 
with generally inhomogeneous mean fields and two-point functions.
Some further discussion may be required to understand the motivation for this
approach.
To start, we note that there is a class of density operators $\hat\r$ which 
can be written as a superposition of gaussian pure states (see
%\cite{MaWo91}
\cite{MaWo91_1,MaWo91_2}
and the appendix A of \cite{SaSm00}):\footnote{Operators are indicated with a
caret.}
\be
   \hat{\r} = \int [d\f\, d\p]\, p[\f,\p]\, |\f,\p\rangle\langle\f,\p|.
   \label{rhtocoh}
\ee
The $|\f,\p\rangle$ are coherent states centered around
$\f(\vecx)= \langle\f,\p|\hat\f(\vecx)|\f,\p\rangle$ and
$\p(\vecx)= \langle\f,\p|\hat\p(\vecx)|\f,\p\rangle$,
and $p[\f,\p]$ is a functional representing the density operator $\hat\r$.

For example, for a free scalar field the canonical distribution
$\hat\r = \exp(-\beta \hat H[\f,\p])$ is represented as (see the appendix A of
\cite{SaSm00} for a derivation)
\be
   p[\f,\p] \propto \prod_{\veck} \exp\left[-(e^{\beta \o_{\veck}}-1)
   (\p_{\veck}^2 + \o_{\veck}^2\f_{\veck}^2)/2\o_{\veck}\right],
   \label{BEinitial}
\ee
where $\veck$ labels the modes of the field with frequency $\o_{\veck}$.

By writing the density operator in this form, and we emphasize that e.g.\ for
the canonical distribution (\ref{BEinitial}) there are no approximations 
involved, we have achieved four things.
Firstly, we have made contact with the classical approximation.
If the mean field in a coherent state is large compared to the width of the
state, the gaussian wavepacket approximately follows a classical trajectory
and the mean field can be thought of as a {\em classical} field. 
This then suggests that the individual coherent states in the ensemble 
may be referred to as ``realizations''.
However, by using an ensemble of coherent 
states rather than classical fields, we have a (hopefully much) better 
description for those modes that have low occupation numbers for which 
the classical dynamics is a poor approximation.
Secondly, we have expressed a (typically non-gaussian) initial density operator
in terms of gaussian states. These are optimal for the Hartree method,
which we want to use to approximate the dynamics of these states.
Thirdly, the mean fields in the individual coherent states are 
{\em inhomogeneous}, therefore the particles can
interact with the inhomogeneous mean field, such that
the energy may get distributed over the full momentum range.
In \cite{SaSm00} we found that this leads to approximate thermalization in
coarse grained distributions. Averaging over the initial ensemble is not
necessary {\em per se}, since it also occurs in each individual member of the
ensemble, because of coarsening over the volume, provided the volume is large
enough to contain a sufficient number of decorrelated systems.
Finally, there is another aspect which is relevant in this context. When
non-perturbative field configurations 
(domain walls, skyrmions, sphalerons, etc.) play a role, these can be
taken into account with inhomogeneous background fields (i.e.\ mean field
realizations). 
This may also be important for thermalization.

In our previous work \cite{SaSm00} the initial state was such that only
a few of the low momenta modes of the mean field realizations
carried all the energy.
Such initial conditions were useful
for equilibration tests. We found partial
thermalization to an approximate Bose-Einstein (BE) 
distribution, for a limited time. 
Gradually the distribution also started to show
classical equipartition features. 
For practical applications it is therefore important to get 
more information on the relevant time scales, 
and this is the main subject of this paper.

In the present work the initial energy is distributed 
over all momentum modes of the realizations as in (\ref{BEinitial}).
This initial density operator is still out of equilibrium because 
of interactions.
As will be shown in the next sections, this leads to 
creation of quantum particles of all energies with a
Bose-Einstein distribution. Depending on the interaction strength
and temperature, this equilibrium may last 
for a long time before classical features begin to dominate, and we are
able to better quantify the relevant time scales.

An important diagnostic in this and our previous 
work is the particle distribution function, which is
defined in terms of equal-time two-point functions. The definition assumes
a quasi-particle picture 
and we check this here by performing a Monte
Carlo computation in imaginary time to compute the exact 
two-point correlation function.

The computations of the quantum mode dynamics is numerically very expensive.
Therefore we also study the effect of reducing the number of quantum 
field modes. With the present initial conditions in terms of a temperature
$T = \beta^{-1}$ it is natural to try eliminating modes with 
$|\veck|\gg T$. 
The often used classical approximation corresponds to the extreme  
of leaving out all quantum modes and to see how close this can mimic the 
quantum world we also make a comparison with this case. 
It turns out that even with a substantially reduced number of modes,
this extended Hartree method fares much better than the classical approximation.

We again use the 
$\f^4$ model in $1+1$ dimension as a test case. This model, the
Hartree ensemble approach and initial conditions
are briefly recalled in Sect.~\ref{method}. In Sect.~\ref{mc} we recall
our definition of the particle distribution in terms of equal time
correlation functions and perform a Monte Carlo check on the underlying
quasi-particle picture.
In Sects.~\ref{wc} and \ref{sc} we study equilibration time scales at weak and
stronger coupling.  
Next, in Sect.~\ref{reduced}, 
we show that the expensive computations of the quantum 
mode dynamics
can be substantially reduced by using only a limited number of mode functions.
In Sect.~\ref{classical} we make a comparison with classical dynamics.  
Finally in Sect.~\ref{discussion} we discuss the results.

\section{Method}
\label{method}

\subsection{ Hartree ensemble approximation}

Consider the Hamiltonian of a scalar field in one dimension,
discretized on a lattice,
\be
   \hat{H} = \sum_x 
   [ \half \ph_x^2 - \half \fh_x (\D \fh)_x + \half \m^2 \fh_x^2 +
          \quart \l \fh_x^4 ],
    \label{HAM}
\ee
with $x=a, 2a, \dots, Na$, $\D\fh_x = (\fh_{x+a} + \fh_{x-a} - 2\fh_x)/a^2$.
The volume is $L=Na$ with periodic boundary conditions; 
The Heisenberg equations follow as,
\be
  \dot{\fh}_x = \hat{\p}_x,\Sp\Sp \dot{\hat{\p}}_x = (\D \fh)_x - \m^2 \fh_x
      -\l \fh_x^3.    \label{HEISENBERG}
\ee

Rather than solving the exact operators from the Heisenberg equation, we
use the gaussian or Hartree approximation (see e.g.\ \cite{SaSm00} for
details). This amounts to approximating the field operators as linear
combinations of time-\emph{in}dependent creation and annihilation operators
$\hat{b}^{\dagger}_\a$ and $\hat{b}_\a$,
\bea
\fh_x & = & \f_x + \sum_\a[ f^\a_x\, \hat{b}_\a + 
f^{\a *}_x\, \hat{b}^{\dagger}_\a],
                \nonumber \\
\hat{\p}_x  & = & \p_x + \sum_\a[ \dot{f}^\a_x\, \hat{b}_\a
                                     + \dot{f}^{\a *}_x\, \hat{b}^{\dagger}_\a],
\label{GAUSSIAN}
\eea
with time-dependent mean fields $\f$ and $\p$, and mode functions $f^\a$. 
All information is contained in the one- and two-point functions, which
can be expressed as
\be
 \av{ \fh_x } = \f_x, \Sp
 C_{xy} = \av{ \fh_x \fh_y } - \av{\fh_x}\av{\fh_y} =  
\sum_\a [ (1+n_\a^0)f_x^\a f_y^{\a *} + n_\a^0 f_y^\a f_x^{\a *}],
\label{CORRFUNC}
\ee
and similarly for the one- and two-point functions involving $\hat\p$.
The $n_\a^0$ are the particle number densities in the initial state.
In our numerical work we shall take $n_\a^0=0$, such that the
system is described by a pure state wave-functional.
All 1PI higher-point functions are zero (i.e. higher-point functions factorize
in one- and two-point functions).
The mode functions represent the width of the wave-functional, allowing
for quantum fluctuations around the mean field. Alternatively one can think
of them as describing the particles in the model.
Since the Hartree method uses (gaussian) wave-functionals, we
expect to improve on the classical dynamics.
Of course we cannot expect to capture all quantum effects, e.g.
tunneling is beyond the scope of this gaussian approximation.

Substituting the gaussian ansatz (\ref{GAUSSIAN}) in the Heisenberg 
equations (\ref{HEISENBERG}) and taking expectation values, we find
self-consistent equations for the mean field $\f_x$ and the mode functions
$f^\a_x$,
\bea
   \ddot{\f}_x & = & \D\f_x - \Big[\m^2 + \hphantom{3}\l \f_x^2 +
             3\l\sum_{\a} (2n_\a^0 + 1)|f_x^{\a}|^2 \Big]\f_x, 
\nonumber \\
   \ddot{f}_x^{\a} & = & \D f_x^{\a} - \Big[ \m^2 + 3\l \f_x^2 +
      3\l\sum_{\a} (2n_\a^0 + 1)|f_x^{\a}|^2 \Big] f_x^\a.
                \label{HARTREE}
\eea
To solve these equations we use a leap-frog algorithm, which is stable for
sufficiently small time steps. Since there are $N$ mode functions (the lattice
has $N$ sites), the amount of work to progress the fields over one time-step is
$O(N^2)$.

Above we described the Hartree approximation. In
the Hartree ensemble method this approximation is applied to
each individual realization $|\f,\p\rangle\langle\f,\p|$ of the
initial conditions as in (\ref{rhtocoh}). So in eq.~(\ref{CORRFUNC}) the
gaussian brackets stand for $\langle \cdot \rangle =
\langle\f,\p|\cdot|\f,\p\rangle$ and the average over $\f,\p$ is only taken in
the evaluation of observables. Furthermore these states are pure, hence
the initial particle density $n_\a^0=0$ in (\ref{CORRFUNC}) and (\ref{HARTREE}).

In this way we compute correlation functions with a generally non-gaussian
density operator 
$\hat\r = \sum_i p_i \ket{\f^{(i)},\p^{(i)} }\bra{\f^{(i)},\p^{(i)} }$, as
\be
  S_{xy} =
   \sum_i p_i [  C_{xy}^{(i)} + \f_x^{(i)} \f_y^{(i)}] 
    -  \Big(\sum_i p_i \f_x^{(i)}\Big)\Big(\sum_j p_j \f_y^{(j)}\Big).
\label{FULLC}
\ee
The $C_{xy}^{(i)}$ and $\f_x^{(i)}$ are computed with gaussian pure
states using (\ref{CORRFUNC}).
This means that in the time-evolution the gaussian approximation is used,
while expectation values are calculated using the more general initial density
operator.

It should be stressed that for typical realizations the mean field
$\f^{(i)}_x$ is {\it inhomogeneous} in space, in contrast to the 
ensemble average $\sum_i p_i \f_x^{(i)}$ which is in fact homogeneous for
the initial conditions we shall employ.
We also note that the equations (\ref{HARTREE}) can be derived from a 
hamiltonian.
Since the equations are also strongly non-linear, this might suggest that the
system could evolve to an equilibrium distribution with equipartition of 
energy, as in classical statistical physics. On the other hand, the mode
functions satisfy Klein-Gordon type orthogonality and completeness relations
that could obstruct such an equipartition and it is not easy
to predict the equilibrium distribution of the model \cite{SaSm00}.

\subsection{ Initial conditions }

In order to solve the equations of motion (\ref{HARTREE}), we must specify
initial conditions for the mean field 
$\f^{(i)}$ and the modes $f^{\a(i)}$ of the individual Hartree trajectories
as well as the weights $p_i$.
This amounts to specifying the initial density operator $\hat\r$.
As explained in the introduction, we use coherent (pure) states to represent
the initial density operator, hence we use initial modes functions as in the 
vacuum state (with the initial particle density $n_\a^0$ equal to zero).
The initial mode functions may be taken as plane waves with wave vector $k$
($\a \to k$), with a normalization appropriate for
a zero temperature free field, 
\be
  f^k_x(0)=\frac{e^{ikx}}{\sqrt{2\o_kL}}, \Sp 
  \dot{f}^k_x(0) = -i\o_k \frac{e^{ikx}}{\sqrt{2\o_kL}},
\ee
(the same for each realization $i$).
$L$ is the system size and $\o_k = \sqrt{m^2 + [2-2 \cos (ak)]/a^2}$,
with $m$ the zero temperature mass.
We will not choose the initial $\f$ 
as far out of equilibrium as we did in ref.~\cite{SaSm00}. There
we took a superposition of only a few low momenta modes, 
\be
\f_x^{(i)} = v, 
\Sp \p_x^{(i)} = A m\sum_{j=1}^{j_{\rm max}}\cos( 2\pi jx/L - \psi_j^{(i)} ), 
\label{INIPAR}
\ee
where $v = \sqrt{m^2/2\l}$ is the vacuum expectation value of the mean field in
the ``broken symmetry phase'' at zero temperature and $\psi_j^{(i)}$ is a random
phase.
Here we choose the mean fields from an 
ensemble with a Bose-Einstein distribution for the 
$\f$ and $\p$ momentum modes as in (\ref{BEinitial}),
\be
p_k (\f_k, \p_k) \propto 
\exp[-(e^{\o_k/T_0}-1)(\p_k^2 + \o_k^2 (\f_k-\delta_{k0} v)^2)/2\o_k ].
                \label{INIBE}
\ee
Then the initial density operator is that of a free field thermal {\em quantum} 
ensemble (cf.\ appendix A in \cite{SaSm00}),
\be
   \hat{\r} = \prod_k \int d\f_k d\p_k\; p_k (\f_k, \p_k) 
   |\f_k,\p_k \rangle \langle \f_k,\p_k|
   \propto e^{-\hat{H}_0/T_0}.
\label{THERMAL}
\ee

It should be stressed that this ensemble is not in equilibrium, 
even though the particle densities we compute from the initial conditions 
(after averaging over a large number of realizations) have a BE distribution.
This is clear, because the mode functions do not contribute at all to the 
initial particle density. In each individual run we therefore expect
quick excitation of the mode functions from their vacuum state,
i.e.\ quantum particles will be created, using energy from the mean field.
Moreover we use the free field dispersion relation
$\o_k^2 = m^2 + [2 - 2\cos(ak)]/a^2$ in the initial distribution 
(\ref{INIBE}), with the zero temperature mass $m \equiv m(0)$.
In thermal equilibrium this should become the temperature
dependent mass $m(T)$ of the quasi-particles. 
Nonetheless we expect that these initial conditions will lead to a much 
faster thermalization than initial conditions of the form (\ref{INIPAR}).

\section{Monte Carlo check}
\label{mc}

To study time-dependence of particle energies and densities 
we define energies $\o_k$ and densities $n_k$ from the Fourier components of
the $\fh$ and $\ph$ two-point functions,
\bea
\frac{1}{L}\sum_{xy} e^{-ik(x-y)} S_{xy}
        &=& (n_k + \half)/\o_k, \nonumber \\
\frac{1}{L}\sum_{xy} e^{-ik(x-y)} U_{xy}
        &=&  (n_k + \half) \o_k,
     \label{DEFNOM}
\eea
where $U_{xy}$ is similarly defined as $S_{xy}$ in (\ref{FULLC}), with $\f$ 
replaced by $\p$.
For weakly coupled fields one expects the equilibrium particle densities 
defined in this way to have a Bose-Einstein distribution, and energies to
have a free quasi-particle dispersion relation, approximately,
\be
  n_k = (e^{\o_k/T} - 1)^{-1},\;\;\;\o_k^2 = m(T)^2 + [2-2\cos(ak)]/a^2.
                       \label{FREEFORM}
\ee
The effective mass $m(T)$ of the quasi-particles is temperature dependent. In
the following we shall use the zero temperature mass $m \equiv m(T=0)$ to scale
dimensionful quantities.

%%%%%%%%%%%%%%%%%%%%  fig 1 %%%%%%%%%%%%%%%%%%%%%%%%%%%%%%%%%%%%
\begin{figure} [tb]
\bc
\scalebox{0.75}[0.85]{ \includegraphics[clip]{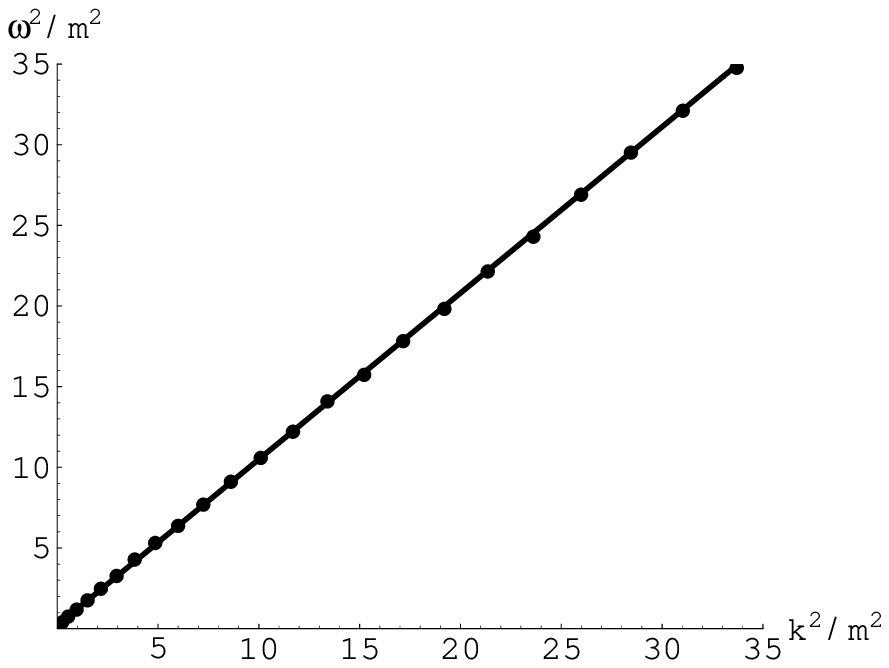} }
\scalebox{0.75}[0.85]{ \includegraphics[clip]{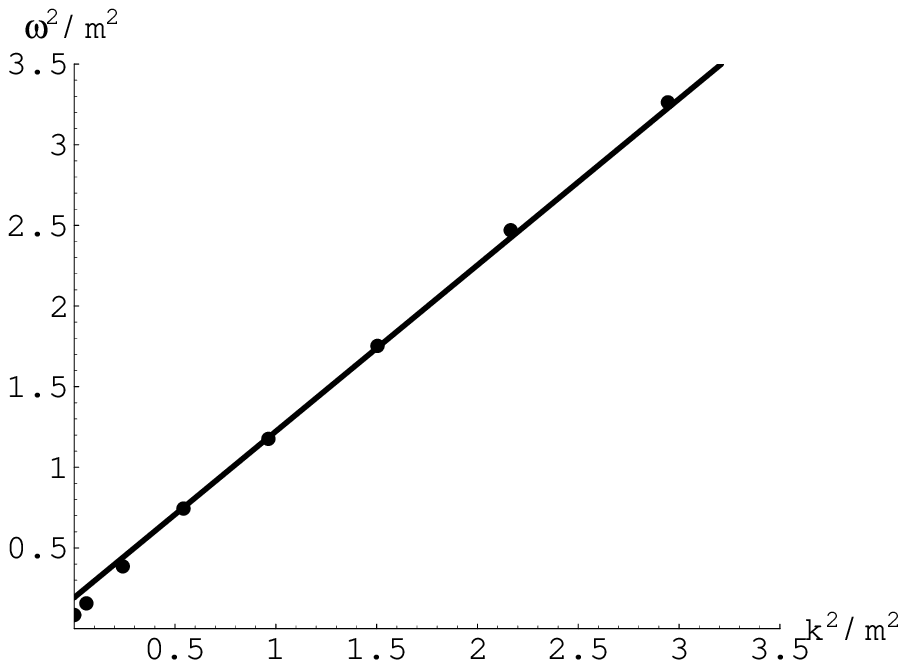} }
\capit{
Dispersion relation computed from a Monte Carlo
simulation of the euclidean time version of the model.
The model parameters are: $\l/m^2 = 1/2v^2 = 1/4$, $Lm = 25.6$,
$1/am = 10$ and $T/m = 1$, with $20$ steps in the euclidean time direction.
Here $k$ is the lattice momentum $\sqrt{2 - 2\cos(ak)}/a$.
The statistical error bars are smaller than the symbols.
The right figure is a zoom-in at small $k$  where deviations from linear
behavior are visible.
}
\label{FIGmc}
\ec
\end{figure}
%%%%%%%%%%%%%%%%%%%%%%%%%%%%%%%%%%%%%%%%%%%%%%%%%%%%%%%%%%%%%%%%

To substantiate this expectation (\ref{FREEFORM}), we have performed several
Monte Carlo simulations of the euclidean time version of our model at parameter
values in the same range as we use for the Hartree simulations. 
In 
Fig.~\ref{FIGmc} we show the dispersion relation 
computed from such a Monte Carlo simulation. 
We chose a temperature $T/m = 1$ and measured  $S_{xy}$.
We stress that such a Monte Carlo simulation gives the {\em exact} 
(up to statistical errors) results for the finite temperature Green function.
Making the assumption that $n_k$ has the BE form, we computed the $\o_k$
from $S_{xy}$ using (\ref{DEFNOM}).
As can be seen from the figure, the free form (\ref{FREEFORM}) for the
quasi-particle dispersion relation holds very well, 
with $m(T)/m \approx 0.43$. This value is close to that found with
the effective potential calculations in the Hartree approximation
\cite{SaSm00}, which gives $m(T)/m \approx 0.41$ at $T/m=1$ 
(see also Fig.~\ref{FIGmassT}).
The effects of the temperature and interactions show up almost exclusively in
the value of the effective mass $m(T)$.

%%%%%%%%%%%%%%%%%%%%  fig 2 %%%%%%%%%%%%%%%%%%%%%%%%%%%%%%%%%%%%
\begin{figure} [tb]
\scalebox{0.88}{ \includegraphics[clip]{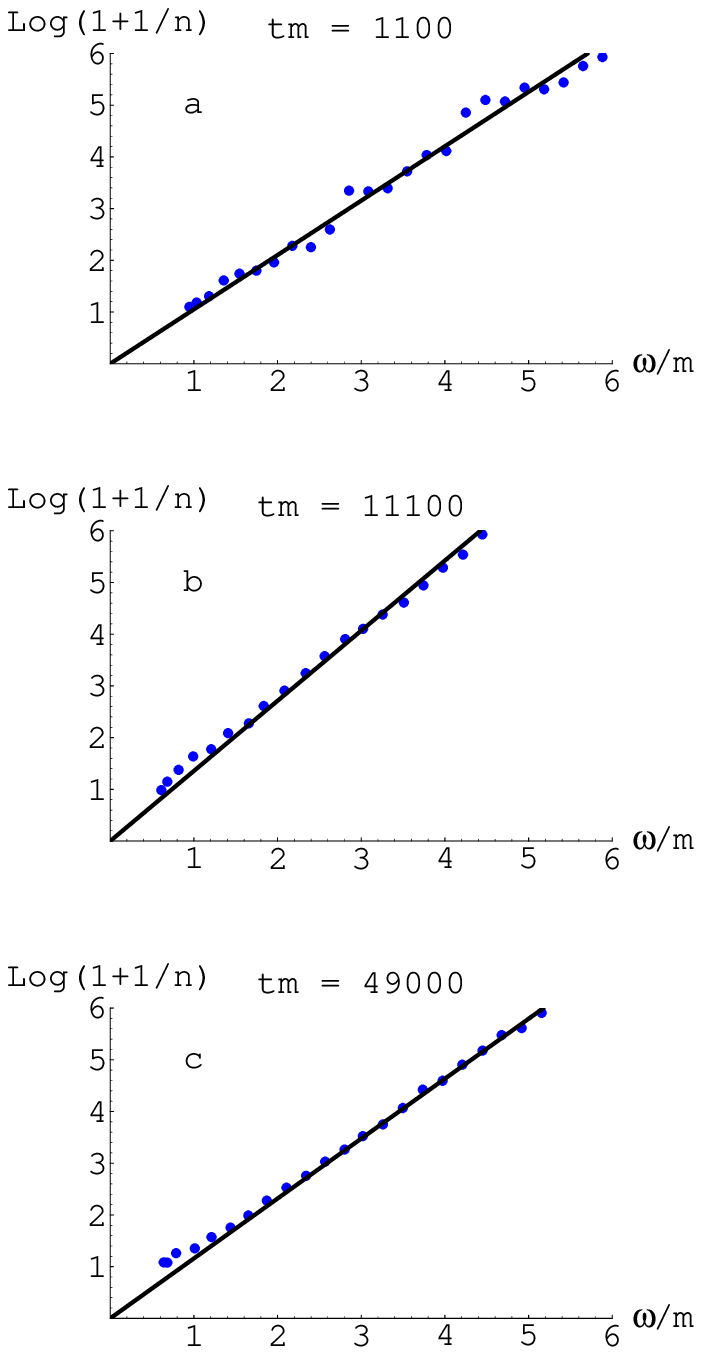} }
\scalebox{0.88}{ \includegraphics[clip]{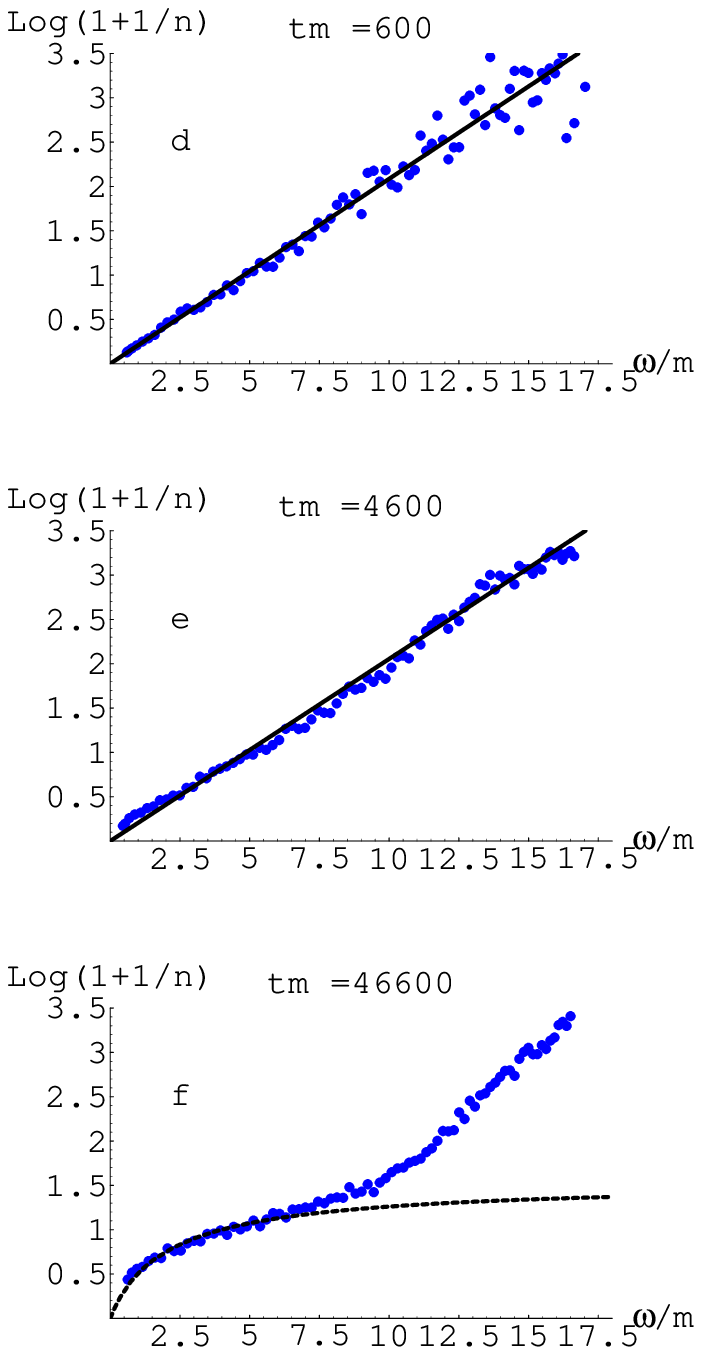} }
\capit{ 
Particle densities as a function of energy, 
plotted as $\log(1 + 1/n)$.
In Figs.~a-c on the left the initial $T_0/m=1$; on the right
the initial temperature is high, $T_0/m=5$.
The model parameters are: 
$\l/m^2 = 1/2v^2 = 1/12$, $Lm = 25.6$, $1/am = 10$. 
}
\label{FIGweak}
\end{figure}
%%%%%%%%%%%%%%%%%%%%%%%%%%%%%%%%%%%%%%%%%%%%%%%%%%%%%%%%%%%%%%%%

\section{Weak coupling}
\label{wc}

In our previous work, using the far out-of-equilibrium initial conditions 
(\ref{INIPAR}),
we found that particles of increasingly higher energy are created and
acquire densities with a BE distribution. However, this thermalization
progressed rather slowly to high energies, 
such that the low momentum particle densities 
already started to deviate from a BE distribution before particles with
energies of a few times the temperature could participate in the equilibrium.
These two phenomena -- particles being created with densities that have
a BE distribution and the gradual emerging of equipartition-like features
 -- will be investigated below using 
the thermal initial conditions (\ref{THERMAL}). 

To probe
the large time behavior we shall use stronger coupling
and higher energy densities than in our previous work.
However, first we show results at the same coupling 
as used in \cite{SaSm00}.
The coupling constant $\l/m^2 = 1/2v^2 = 1/12$ is
in the ``broken symmetry phase'' of the model. The volume $Lm=25.6$
and the lattice cut-off $1/am = 10$.

We plot $\log(1 + 1/n_k)$ rather than $n_k$ itself, because in this way
a BE distribution shows up as a straight line with a slope equal to the
inverse temperature. 
The scattering in the data points is due to using only a few
Hartree realizations (only two initial conditions).
At low temperature, $T_0/m=1$, Figs.~\ref{FIGweak}a-c, the evolution
is very slow and there is hardly a sign of emerging classical features even
at the largest time $tm\approx 50000$. 
Even though the particle distribution does not 
change, there is a persistent, slow transfer
of energy from the mean field into the modes. At $tm=200$, 50\% of the energy
is still in the mean field, at $tm=6000$ this has dropped to 25\% and at
$tm=50000$ it is still some 15\%. The effective mass stays roughly constant,
$m(T)/m \approx 0.94$, which is consistent with the effective potential for
$T_0/m=1$.

At higher temperature, $T_0/m=5$, but with the same weak coupling, there is
again a wide window in which the particles have a BE distribution without
significant distortions (Figs.~\ref{FIGweak}d-f). However, we see classical-like
features emerging for $tm\gtrsim 4000$: compared to the BE distribution, the low
momenta modes become under-occupied, while the high momenta modes become
over-occupied. We find that at the latest time $tm\approx 50000$ the
distribution for $\o/m \lesssim 7$ can be described reasonably well with an
ansatz $n_k=c_0 + T_{cl}/\o_k$. Without the constant $c_0\approx 0.25$ the fit
would be poor.

%%%%%%%%%%%%%%%%%%%%  fig 3 and 4 %%%%%%%%%%%%%%%%%%%%%%%%%%%%%%%%%%%%
\begin{figure} [tb]
\bc
\scalebox{0.87}[0.65]{ \includegraphics[clip]{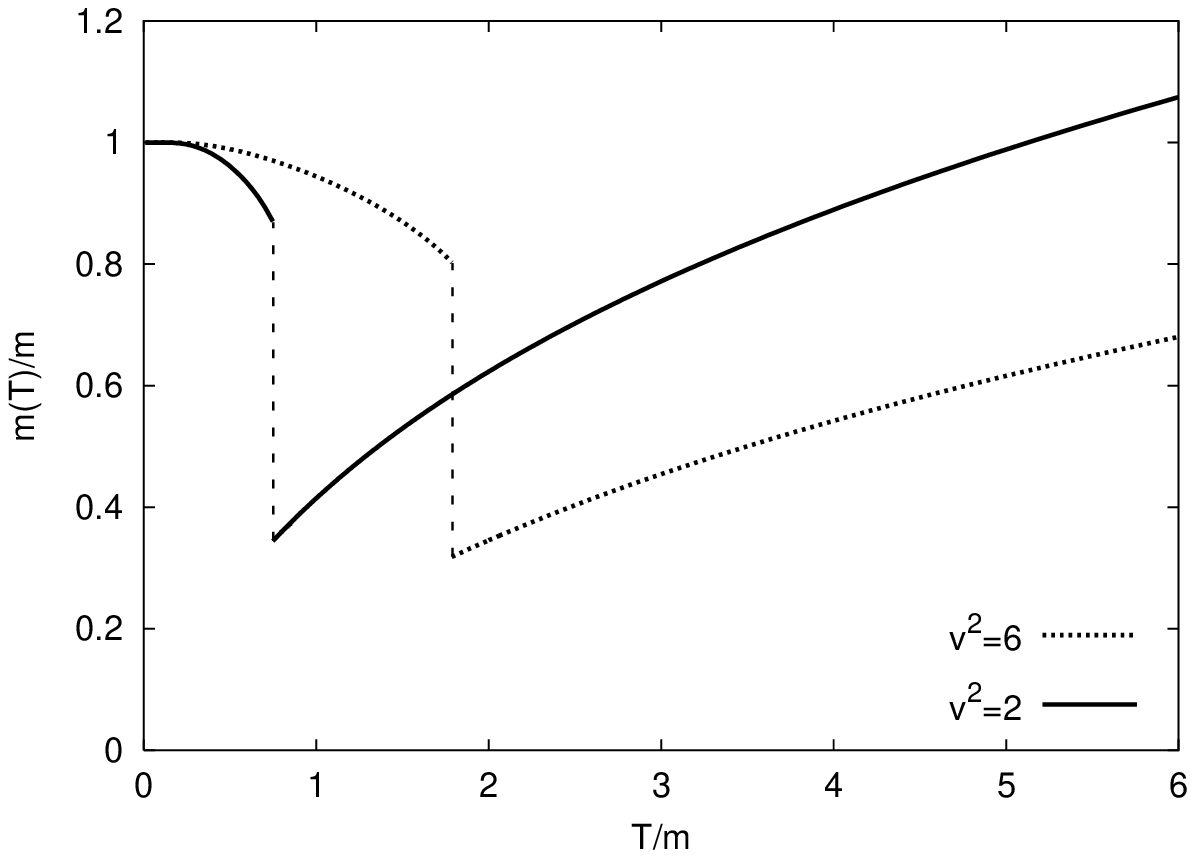} }
\capit{Temperature dependence of the effective mass computed using the
Hartree effective potential \cite{SaSm00}, 
$\l/m^2 = 1/4$ (solid line) and $1/12$ (dotted line), $mL=16$ (volume
dependence is very small).
}
\vspace{0.3cm}
\label{FIGmassT}
\scalebox{1.03}[0.80]{ \includegraphics[clip]{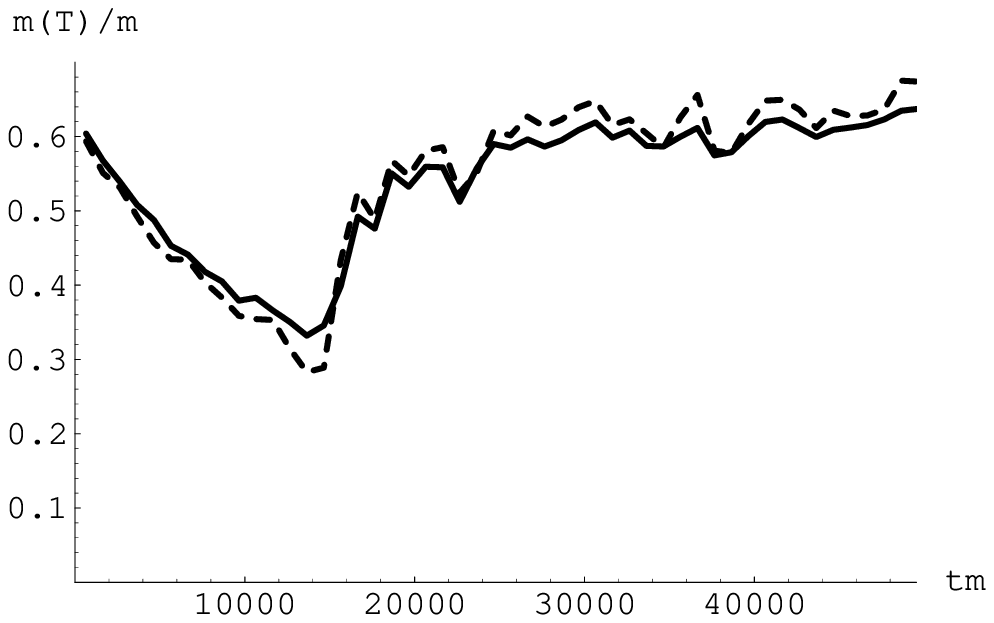} }
\capit{ Time dependence of the effective mass $m(T)$ for the same model
as shown in Figs.~\ref{FIGweak}d-f. The mass is determined as the
lowest energy $\o_0$ (dotted line) or from a quadratic fit to the
dispersion relation (full line).
}
\label{FIGmass}
\ec

\end{figure}
%%%%%%%%%%%%%%%%%%%%%%%%%%%%%%%%%%%%%%%%%%%%%%%%%%%%%%%%%%%%%%%%
\vspace{0.1cm}

In this simulation we find an interesting behavior of the effective mass,
shown in Fig.~\ref{FIGmass}. 
For comparison, we also show in Fig.~\ref{FIGmassT} the effective
mass calculated using the Hartree effective potential at the same model
parameters.  First the mass is steadily
decreasing, which is appropriate when the temperature is decreasing and
the system is in the hot, symmetric phase. At  $tm\approx 14000$ there is
a sharp turnover and the mass starts to increase as in the cold, broken phase.
The temperature at that point $T_{cl}/m \approx 1.6$, 
obtained from a classical fit, is close to the temperature $T_c/m=1.8$ of the 
first order phase transition computed
from the effective potential.\footnote{We recall that in the exact theory
there would be a cross-over instead of a first order phase transition.}
Also the average mean field fluctuates around zero before and around $v\approx
1.8$ after the transition, reasonably close to the effective potential
prediction $v\approx2$ for $T/m\sim1.6-1.8$. The reasonable quantitative
agreement between the simulation, which shows classical features, and the
effective potential computation, which assumes a BE distribution, illustrates
that the thermal mass is dominated by the low-energy particles, for which there
is little difference between a BE and classical distribution.

%%%%%%%%%%%%%%%%%%%%  fig 5 %%%%%%%%%%%%%%%%%%%%%%%%%%%%%%%%%%%%
\begin{figure} [tb]
\scalebox{0.88}{ \includegraphics[clip]{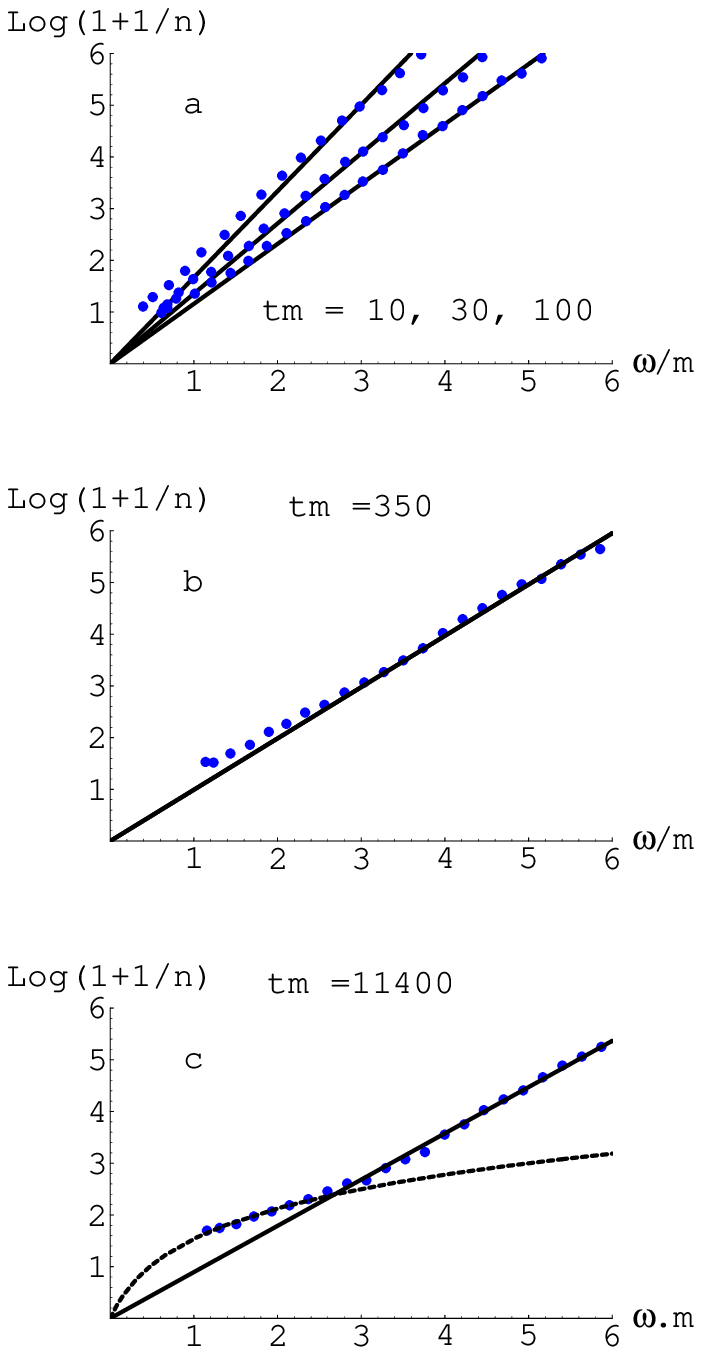} }
\scalebox{0.88}{ \includegraphics[clip]{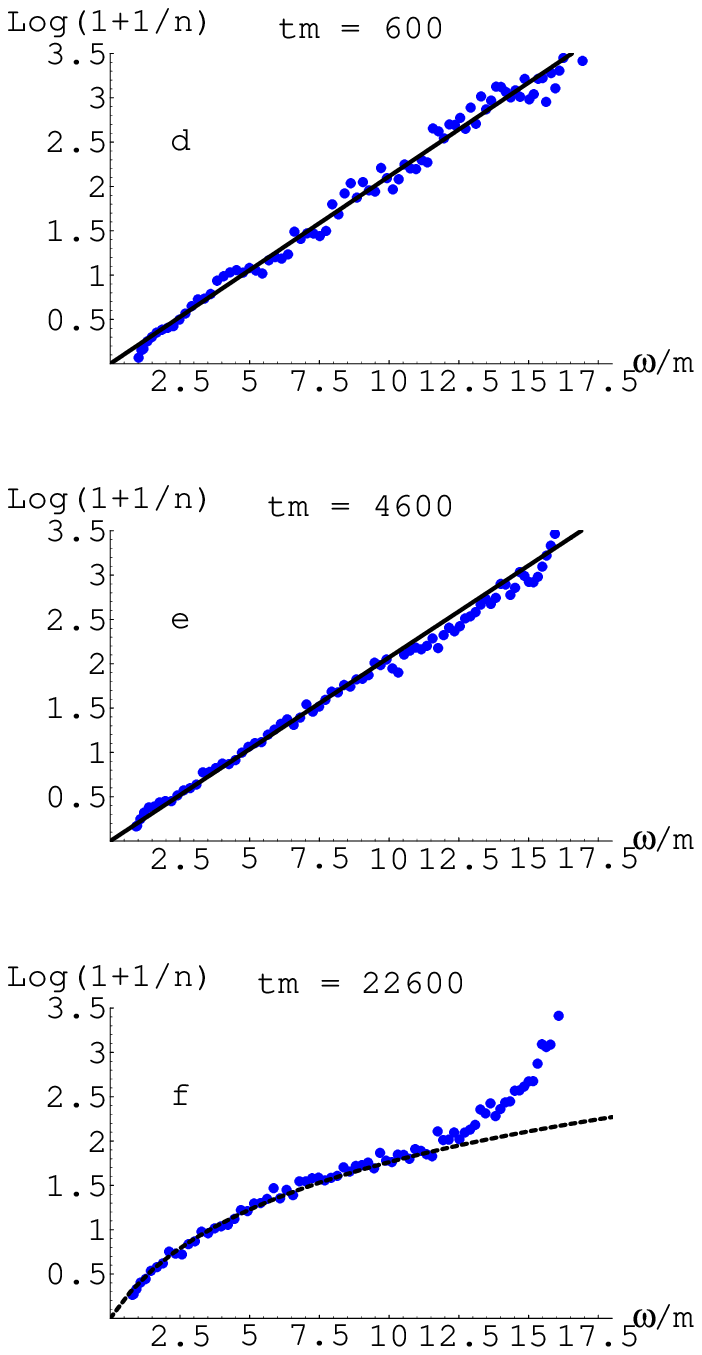} }
\capit{ 
The same as Fig.~\ref{FIGweak}, but at a stronger coupling $\l/m^2 = 1/4$.
}
\label{FIGref}
\end{figure}
%%%%%%%%%%%%%%%%%%%%%%%%%%%%%%%%%%%%%%%%%%%%%%%%%%%%%%%%%%%%%%%%

\section{Stronger coupling}
\label{sc}

We now turn to the stronger coupling $\l/m^2 =1/2v^2= 1/4$, in order to make
processes evolve faster. 
In Fig.~\ref{FIGref}a we show particle densities $n_k$ computed only 
from the mode functions. We ignore the contribution from the mean
field in (\ref{FULLC}), because we want to focus on the particles 
described by the mode functions. 
In Fig.~\ref{FIGref}a one sees that already after a short time,
$tm \gtrsim 10$, 
particles have been created over a wide range of energies,
$\o/m \lesssim 6$. 
The densities are reasonably well described by a BE distribution with
a time dependent temperature. This temperature initially increases
rapidly from $T/m=0$ at $tm=0$ to $T/m\approx 0.6$ at $tm=10$ and then 
gradually increases further to $T/m\approx 0.9$ at $tm=100$. 
(Recall that the temperature of the initial condition is $T_0 = m$.)

Figs.~\ref{FIGref}b-c, which are obtained using both the modes and mean fields
to compute the correlation functions, show that the densities of particles
with large momenta tend to remain at a BE distribution also for later times,
with a very slowly increasing temperature $T/m = 0.93-1.13$. However, one also
clearly sees deviations from the BE distribution developing, starting at the
low $\o$-side of the spectrum.

%%%%%%%%%%%%%%%%%%%%  fig 6 %%%%%%%%%%%%%%%%%%%%%%%%%%%%%%%%%%%%
\begin{figure} [tb]
\scalebox{0.85}[1.05]{ \includegraphics[clip]{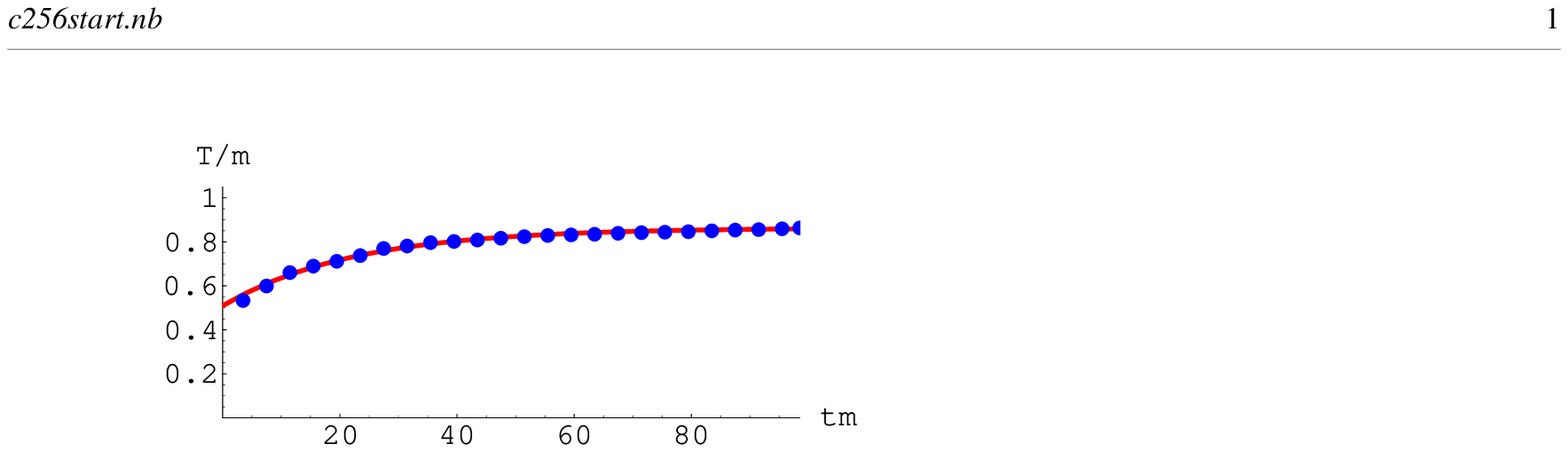} }
\scalebox{0.85}[1.05]{ \includegraphics[clip]{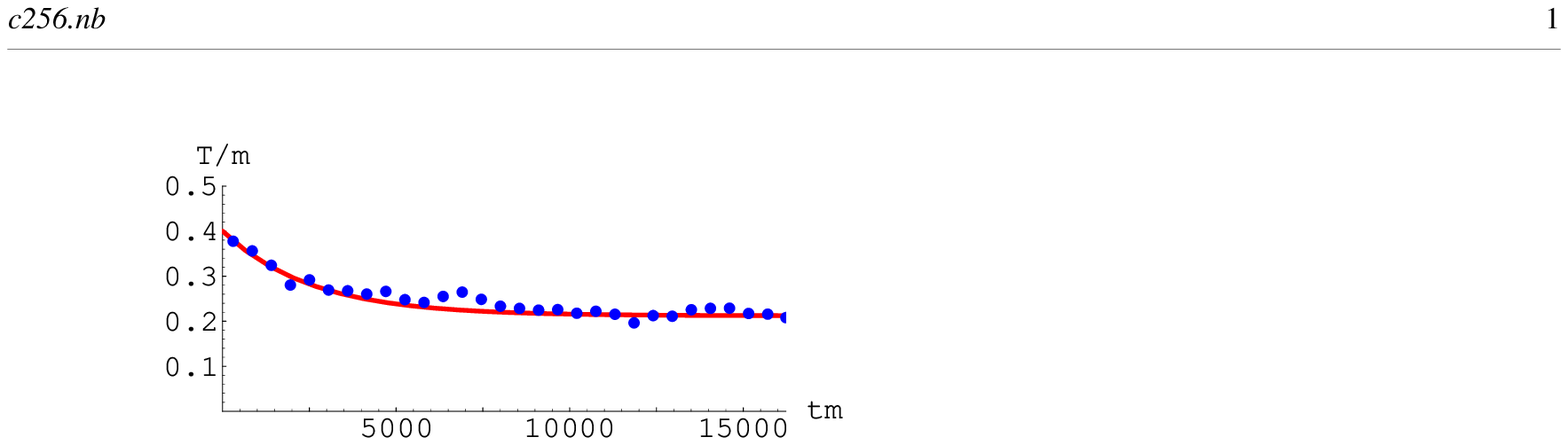} }
\vspace{4cm}
\capit{Time dependence of BE (from the modes only) and classical (from the modes
and mean field) temperatures for the data of Figs.~\ref{FIGref}a-c.
}
\label{FIGtimes}
\end{figure}
%%%%%%%%%%%%%%%%%%%%%%%%%%%%%%%%%%%%%%%%%%%%%%%%%%%%%%%%%%%%%%%%

From these data we infer two time scales: First there is the rate
at which the temperature of the BE distribution of the quantum particles
is established. Second there is a rate at which the classical-like distribution
sets in. Fig.~\ref{FIGtimes} shows the time dependence of these two
processes. The BE temperature was computed by fitting $\log(1+1/n) = \o/T$
(only using the mode function contribution) for $2 \lesssim \o/m \lesssim 4$.
The classical temperature was found from fitting $n=T_{cl}/\o$ for 
$\o/m\lesssim 2$. 
The time dependence of these temperatures
is reasonably well described by an exponential approach to an equilibrium
value, $T_{BE}(t) = A - B e^{-t/\t_{BE}}$
and  $T_{cl}(t) = A' + B' e^{-t/\t_{cl}}$. 
We find $m\t_{BE} \approx 20$ and
$m\t_{cl}\approx 2500$, showing quantitatively that the approximate BE
thermalization happens much faster than the emergence of classical-like behavior
(Note that $T_{cl}$ becomes much lower than $T_{BE}$, which itself is somewhat
smaller than $T_0$, in agreement with the eventually expected classical
equipartition).

At higher temperature, the distribution roughly follows the same development.
Surprisingly enough the distribution keeps its approximate BE form much longer,
while at a higher temperature one expects a stronger effective coupling,
and thus shorter time-scales.
In Figs.~\ref{FIGref}d-f the initial temperature is $T_0/m=5$. At this higher
temperature and on a correspondingly larger energy scale, the deviations from a
BE distribution appear less pronounced at early times. But even at $tm=4600$ the
particle densities are reasonably well described by a BE distribution with a
temperature $T/m \approx 4.8$. At this time, there is a small reduction
of the density of low momentum particles ($n$ is up to $15\%$ smaller than the
BE density, which is hard to see on the $\log$-plot). At the same time the
density of particles, with $\o/m$ in the region $10-12$, increases a little.
This trend continues and at $tm=22600$ there is classical behavior for $\o/m
\lesssim 12$. 

The dashed line in Fig.~\ref{FIGref}f is a
fit of the form $n = T_{cl}/\o$, which gives a ``classical'' temperature
$T_{cl}/m \approx 2.1$. The good quality of this fit for $\o/m \lesssim 12$
suggests that the BE distribution gradually turns over into classical
equipartition. However, for still larger times the distribution is no
longer well described by a simple $n \propto 1/\o$-dependence. We did not
determine the final equilibrium distribution, because of the extremely long
(computer) time this would require.

\begin{table}
\begin{center}
\frame{
\begin{tabular}{|l||l|l|l|l|}
\hline
$\l/m^2$  & \multicolumn{2}{c|}{1/12}  &\multicolumn{2}{c|}{1/4} \\
\hline
$T_0/m$   &1  &5 &1 &5\\
\hline
\hline
$m\t_{BE}$ & 35      & 35      & 25       & 25    \\
\hline
$m\t_{cl}$ & $> 15000$ &3000-5000 &2500-3500 &2000-5000 \\
\hline
\end{tabular}}
\caption{{\em
Results for the BE equilibration time, $\t_{BE}$,
and the time scale for the drift towards classical equipartition, 
$\t_{cl}$, obtained from fits to $T_{BE}$
and $T_{cl}$ as in Fig.~\ref{FIGtimes}, as well as similar fits to the
time dependence of $\sum_k n_k$.}}
\end{center}
\end{table}

In Table~1 we summarize our results for $\t_{BE}$ and $\t_{cl}$,
including also fits to the time dependence of the particle density 
$\sum_k n_k$, computed from the modes only or mean field plus modes, as in
Figs.~\ref{FIGtimes}. These results do not show a clear dependence on the 
coupling or temperature, contrary to the expectation of much smaller time 
scales at higher temperature and/or stronger coupling. 
We believe that this is accidental, due to the fact that
at $T_0/m = 5$ and/or $\l/m^2 = 1/4$ the system is in the 
``symmetric phase'', while it is in the ``broken phase'' for $T_0/m = 1$
and $\l/m^2 = 1/12$. We have noticed previously \cite{SaSm00}
that in the ``symmetric phase'' the system evolves much more slowly than in
the ``broken phase''. Note that the numbers in the table are subject to 
systematic uncertainty, since the mode system starts far from equilibrium and 
the time dependence not always follows an unambiguous exponential relaxation.
This applies in particular for the simulation at $\l/m^2=1/4$, $T_0/m=1$,
which is very close to the ``phase transition''.
% Non-equilibrium distributions such as $n_k(t)$ versus $\o_k(t)$
% are sometimes difficult to interpret.

Besides looking at the particle number distribution, it is interesting to
follow the effective mass $m(T)$ in time. Comparing it with the
temperature dependence computed analytically  using the Hartree effective
potential \cite{SaSm00}, gives another measure for the
effective temperature of the system. The simulation of Figs.~\ref{FIGref}a-c
gave an effective mass which increased slightly in the range 
$m(T)/m = 0.84-0.89$.
From the effective potential calculation we then infer that the temperature
should be in the range $0.5\lesssim T/m \lesssim 0.7$, i.e.\ in 
the ``low temperature''
phase of the model, cf. Fig.~\ref{FIGmassT},
which is confirmed by checking the values of the mean
field. This temperature is considerably lower than the BE temperature 
$T/m= 1.0(1)$ estimated from the particle distribution at higher momenta, 
but is consistent with the temperature obtained from fitting 
$n_k$ at the smaller $\o_k$ with a classical distribution.
The same is found for the high-temperature simulation of
Figs.~\ref{FIGref}d-f: $m(T)/m$ decreases from $1.12$ at the start to $0.60$ 
at $tm=22600$,
which corresponds, using the effective potential, to a decrease from
$T/m\approx 5$ to $T/m\approx 2$, consistent with the observed
$T_{cl}/m\approx 5$ to $T_{cl}/m\approx 2.1$.
As mentioned before, the difference between the classical and BE distribution
is unimportant for the dominant $n_k$, those at low momenta.

%%%%%%%%%%%%%%%%%%%%  fig 7 %%%%%%%%%%%%%%%%%%%%%%%%%%%%%%%%%%%%
\begin{figure} [tb]
\scalebox{1.0}{\includegraphics[width=\textwidth,clip]{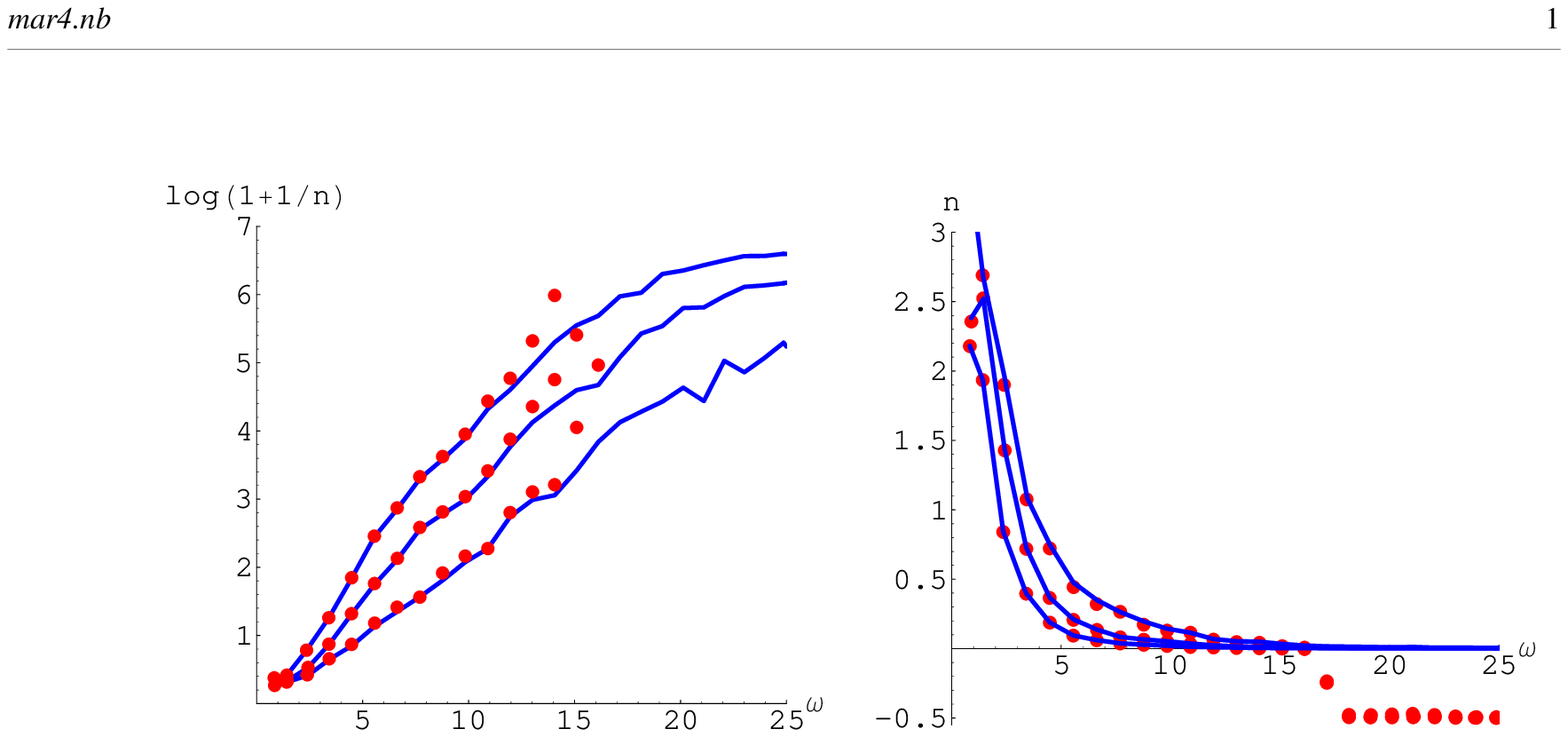}}
\vspace{6cm}
\capit{Particle densities at $tm=50, 90, 300$, obtained from simulations using
the full number of modes (drawn lines) and using only modes for which $\o_k/m <
17 \approx 3 T/m$. The left figure shows $\log(1+1/n)$, the right shows the
density $n$ itself ($Lm=5.7$, $1/am = 22.3$ and $\l/m^2 = 1/12$).
\label{FIGcomp}
}
\end{figure}
%%%%%%%%%%%%%%%%%%%%%%%%%%%%%%%%%%%%%%%%%%%%%%%%%%%%%%%%%%%%%%%%

\section{Reduced number of mode functions}
\label{reduced}

If the positive results for the performance of the
Hartree ensemble method at shorter times 
carry over to more realistic models
in $3+1$ dimensions, one has to confront the problem of the high
computational cost of this approach. Taking the continuum limit
on a finite volume in $d$ dimensions, i.e. increasing the number of
lattice sites $N$ in each direction, the cost of our approach increases
$\propto N^{2d+1}$: There are $O(N^d)$ fields which have to be
updated $O(N)$ times, assuming a fixed value of the time-step $a_0/a$.

Most of this cost comes from having to solve all $N^d$ mode functions
$f^\alpha$.
However, many of these modes would represent 
particles with very high momenta $|k| \gg T$. Such particles have
very low densities and should be irrelevant for the 
physics at lower scales. This suggests reducing the number of mode
functions in our simulations. We have tested this idea by comparing
a simulation on a lattice with $N=128$ sites using all $128$ mode functions,
with the same simulation using only $32$ mode functions. This induces a 
maximum energy $\o_{\rm max}/m \approx \sqrt{2-2\cos(32\p/128)}/am \approx 17$,
which is much larger than the temperature $T/m \approx 6.8$ that we will
use in the test. The $\o_{\rm max}$ here refers to the energy of the initial 
mode functions which are plane waves.

In order to make as detailed a comparison as possible, we show the
results for the particle density obtained from the mode functions only,
leaving out the contribution of the mean field in (\ref{FULLC}).
As shown in Fig.~\ref{FIGcomp}, this partial distribution changes with time 
during thermalization (cf.~Fig.~\ref{FIGref}a).
The drawn lines represent the data obtained with the full number of
mode functions.
The dots represent the results obtained using the reduced number of modes. 
The left figure shows the familiar $\log(1+1/n)$ form 
of the density, the other figure shows the density $n$ itself.
As is evident, these results reproduce the data from
the reference simulation accurately up to $\o/m\approx 12$, which is
close to $\o_{\rm max}/m \approx 17$. Notice that the densities, computed from
eq.~(\ref{DEFNOM}), drop to $n = -0.5$ for $\o/m\gtrsim 17$
(Fig.~\ref{FIGcomp} right). For these high
momenta there are no more mode functions available to provide the 
vacuum fluctuations that should lift the density to zero.
It indicates that at high momenta there is still a roughly one-to-one
correspondence between mode functions and momentum labels of the 
particles.

\section{Classical approximation}
\label{classical}

Even using fewer mode functions, the Hartree 
approach is much more expensive than the classical
approximation (which has no mode functions). 
So it is important to check if our results cannot in some way be mimicked by
a classical approximation. The standard way to implement 
the latter at high temperature, is to average over initial configurations drawn
from the Boltzmann distribution. 
Up to modifications by the interactions
this implies a classical distribution function $n(\o) = T/\o$, with a slow
fall off causing Rayleigh-Jeans type divergences. 
Actually, in $1+1$ dimensions such divergences are absent in $\f\f$-correlation
functions \cite{SmTa99}.

%%%%%%%%%%%%%%%%%%%%  fig 8  %%%%%%%%%%%%%%%%%%%%%%%%%%%%%%%%%%%%
\begin{figure} [tb]
\scalebox{0.88}{ \includegraphics[clip]{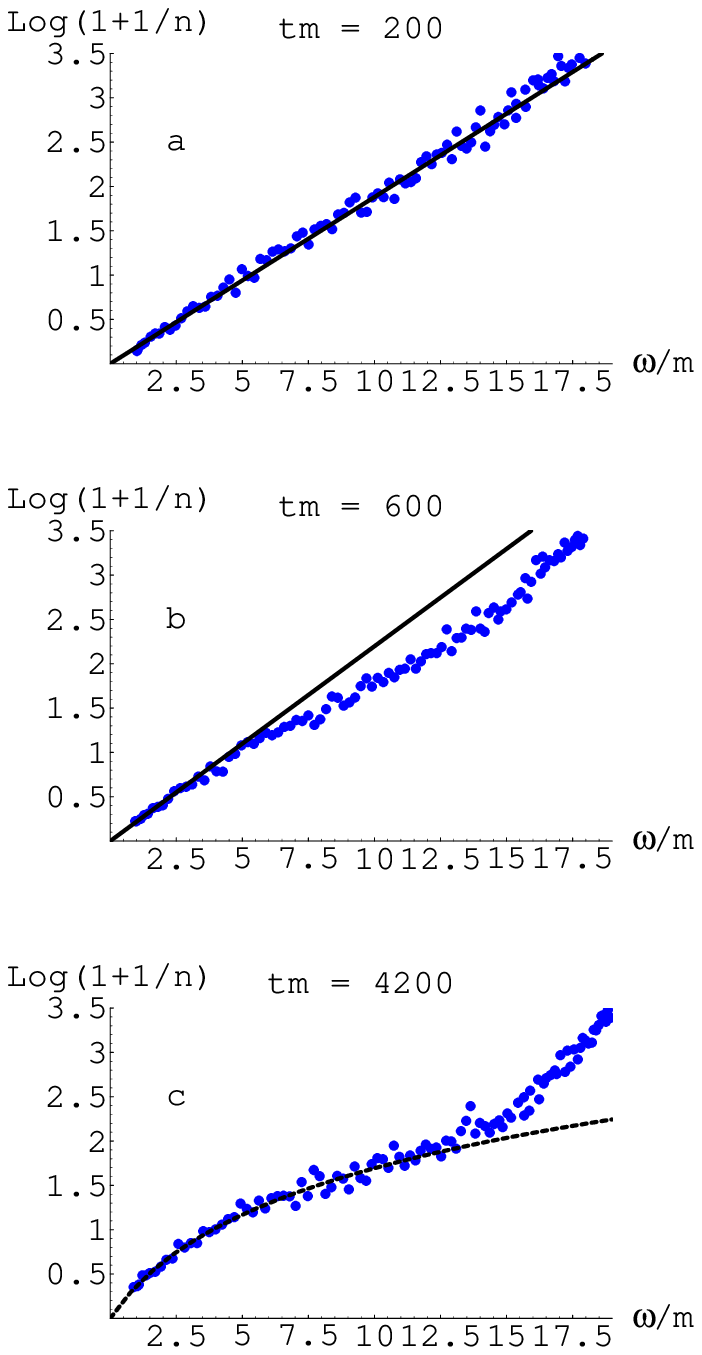} }
\scalebox{0.88}{ \includegraphics[clip]{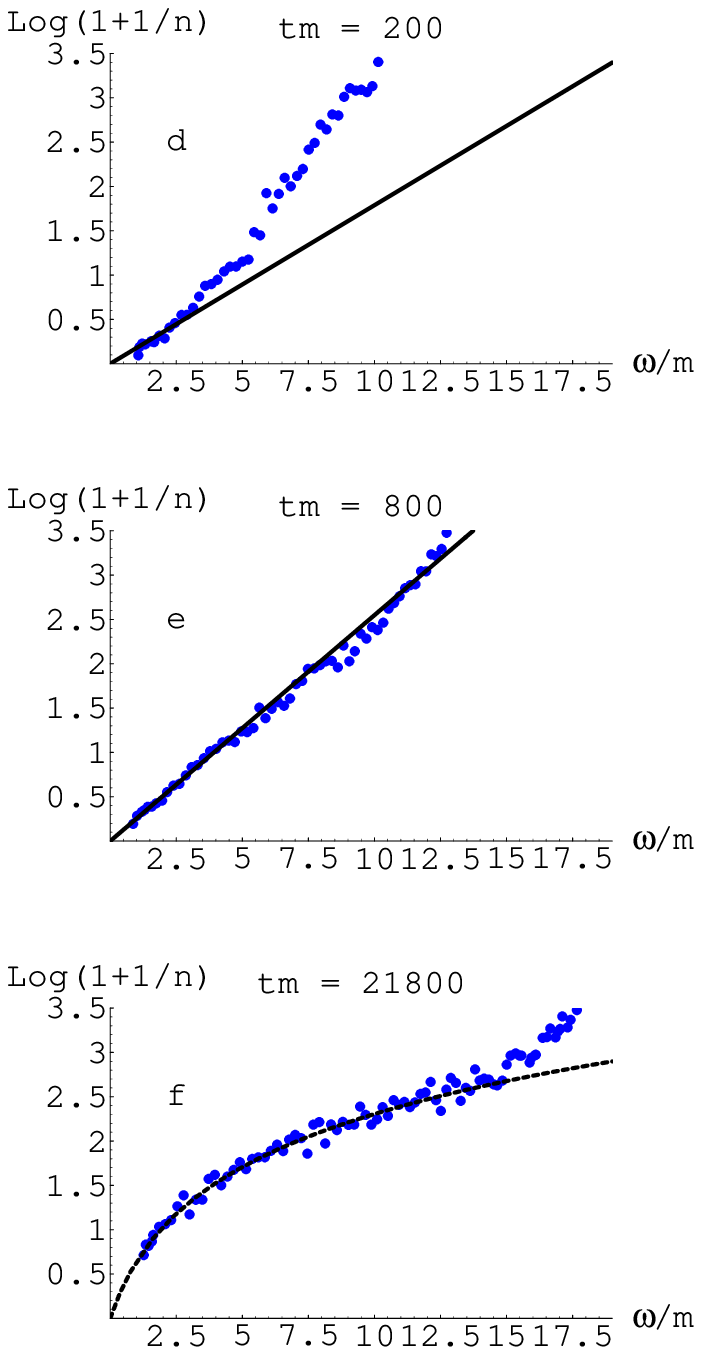} }
\capit{ Same as Figs.~\ref{FIGref}d-f, but using classical dynamics.
Figs.~a-c have BE-type initial conditions with temperature $T_0/m=5$,
in Figs.~d-f only a few of the low momentum modes of the mean field
carried all energy. The other model parameters are:
$\l/m^2 = 1/4$, $Lm = 25.6$ and $1/am = 10$.
}
\label{FIGclass}
\end{figure}

%%%%%%%%%%%%%%%%%%%%%%%%%%%%%%%%%%%%%%%%%%%%%%%%%%%%%%%%%%%%%%%%

Here we want to ask a
somewhat different question: to what extend can classical dynamics
be used to represent a thermalized system with a Bose-Einstein distribution 
for the particle densities? To investigate this we shall use 
the same BE-type initial conditions (\ref{INIBE})
as in the Hartree case,
as well as the much more out-of-equilibrium
conditions of the form (\ref{INIPAR}). 
We perform similar simulations and analyses as
before, but now without mode functions (and with $n_k+1/2 \to n_k$ in
(\ref{DEFNOM}), as there is now no quantum vacuum contribution).
Typically, the classical dynamics produces data with more noise, since the 
average
contribution of the many mode functions tends to smoothen results in the 
Hartree-dynamics. We counter this noise by averaging over 40-50 initial
conditions, which is much more than we typically use with Hartree dynamics.
We note in passing that the necessity to use a larger ensemble to obtain
data of the same quality as with Hartree dynamics, diminishes the computational
advantage of using classical dynamics considerably.

At the same weak coupling and low temperature as in Figs.~\ref{FIGweak}a-c, 
we find that the classical system also 
preserves the Bose-Einstein distribution of the initial conditions very well.
Even at the largest simulated time, $tm = 50000$, there is no compelling
sign of equipartition in the particle distribution. This, however, may only
show that the relaxation time-scale of the classical dynamics is very long,
cf.~\cite{AaBo00}, and that we are seeing remnants of the initial condition
rather than thermalization.

To speed-up the dynamics we increased the temperature to $T_0/m=5$ 
and the coupling to $\l/m^2=1/4$, as was
used in Figs.~\ref{FIGref}d-f. The results are
shown in Figs.~\ref{FIGclass}a-c. Now the
initial BE-distribution still persists for some time,
but already at $tm\approx 600$ there are clear signs of equipartition setting
in, whereas effects of similar magnitude only emerged at $tm \gtrsim 6000$ with 
Hartree dynamics.
The gradual move from the initial 
state towards classical equipartition happens much faster
than in the Hartree ensemble simulations.

Of course one might argue that this initial persisting of the  BE
distribution is of little significance, since
it only demonstrates that it takes time to loose the
effect of the initial conditions. In more realistic models we might
not be able to specify initial conditions sufficiently close to thermal 
equilibrium and then one may not expect to encounter a BE distribution.
Yet, somewhat surprisingly, starting with the far
out-of-equilibrium initial condition (\ref{INIPAR}),
we see that as the model steadily moves towards
classical equipartition, the particles are
distributed in a BE-like way in an intermediate stage. 
This can be seen in Figs.~\ref{FIGclass}d-f, where we show particle density
distributions
at simulation times in the range $tm = 200-20000$.  At $tm$ around $800$
the particle densities follow a BE-like distribution over a wide range of
energies. The coupling strength in this simulation is $\l/m^2 = 1/4$,
at weaker coupling this intermediate stage with BE-like distribution
persists longer. However, it always smoothly turns towards classical
equipartition on much shorter time-scales than when using Hartree dynamics
(although, as mentioned above,
the final equilibration to the classical distribution takes place on a
very long time scale).

\section{Discussion}
\label{discussion}

Using Monte Carlo methods we checked that the quasi-particle assumption,
which is basic to our definition of the particle distribution function,
is well satisfied in the range of couplings and temperatures considered here
and in \cite{SaSm00}. 

From the results in this work combined with \cite{SaSm00}, 
the following picture has emerged 
for the 1+1 dimensional $\lambda\varphi^4$ model in the ``broken phase''. 
The initial energy, which is put solely in the mean field of a realization,
is subsequently transfered to the mode functions. This process takes place 
fairly locally in momentum space, i.e.\ mean field modes with momentum $k$ 
excite primarily particle modes with momenta close to $k$, and the modes
then thermalize locally to a BE distribution. In our previous work this
approximate thermalization was more conspicuous because the initial distribution
was further out of equilibrium. Here the BE distribution was put into the
initial condition for the mean fields. However, the corresponding density 
operator
is still out of equilibrium because of the ``wrong'' initial thermal mass. The
thermalization process is fairly rapid, 
within a time $\t_{BE}\approx 25 - 35\, m^{-1}$, 
for $\l/m^2=1/4$, 1/12 and $T_0/m = 5,1$, 
as determined from the time-dependence of the BE temperature 
%(as in Fig.~\ref{FIGtimes}, left)
or the particle density $\sum_k n_k$ (cf.\ Table 1).

This time scale is similar to our findings with initial mean fields containing
only low momenta \cite{SaSm00}. The subsequent thermalization of higher momenta
is very slow. We ascribe this to a weakening of the non-linearities when the
mean field looses much of its energy. When the mean field fluctuates around its
(temperature dependent) equilibrium value, with diminishing amplitude, the
dynamics becomes approximately that of Hartree with a homogeneous mean field,
suggesting lack of thermalization. 
This also explains why the evolution to a classical-like distribution is
much slower with the Hartree ensemble approximation than using classical
dynamics.

However, the fluctuations die out very slowly
and even at very large times of order $10^4\, m^{-1}$ there is still $O(10\%)$
of the energy in the fluctuating mean field. Nonlinear fluctuations remain,
which lead eventually to classical-like equipartition (according to the
effective hamiltonian and conserved ``charges'' \cite{SaSm00}).

The time scale for such classical equipartition setting in could not be
determined in \cite{SaSm00}, its determination is one of the results
of the present work.
%If the coupling is sufficiently weak and/or the temperature sufficiently
%high (as in Figs.~\ref{FIGref}d-f and Fig.~\ref{FIGweak}), we find that 
%the system remains in an approximate quantal thermal state for times of 
%the order $\t_{cl} = 10-100\, \t_{BE}$. 
We find that the system remains in an approximate quantal thermal state for 
times of the order $\t_{cl} \gtrsim 100\, \t_{BE}$ (cf.\ Table 1).

This is an encouraging result.
For example, in a crude application of our 1+1 dimensional results to 3+1
dimensional heavy ion collisions, identifying $m$ with the mass of the
$\s$-resonance $m_{\s}= 600-1200$ MeV, say 900 MeV, 
a time-span of $100\, \t_{BE} = 2000 \, m^{-1}$ 
would correspond to a reasonable length of about 450 Fermi. Within such a
time-span the Hartree ensemble method may be a definite improvement on the
classical dynamics usually employed for e.g.\ the ``disoriented chiral
condensate''.

For application to 3+1 dimensions it is important 
that the numerical efficiency of the Hartree 
method can be significantly improved by using only
a limited number of mode functions, 
corresponding to particles with sufficiently high densities 
(see sect.~\ref{reduced}).
 
Leaving out the mode functions altogether, 
i.e.\ using classical dynamics, 
the results were qualitatively similar to 
those with Hartree dynamics, but 
the emergence of classical particle 
distributions goes faster by roughly an order of magnitude.
So this may not be good enough for practical applications.

With respect to thermalization it is good to keep in mind that 
in the Boltzmann approximation, the collision term corresponding to $2-2$
scattering is identically zero, due to kinematical constraints
in the $\f^4$ model in 1+1 dimensions. So thermalization has to come from
inelastic scattering and/or off-shell effects.
It is then important to realize that such effects are more pronounced in
the ``broken phase'' of the model, which has a three point vertex and finite
(as opposed to zero) range interactions. As mentioned in \cite{SaSm00},
thermalization is drastically less efficient in the ``symmetric phase'' at
similar values of $\l/m^2$. It is sobering to recall the huge thermalization
times found in \cite{AaBo00} in the classical approximation, in the ``symmetric
phase''.
For example, using $\l/m^2 = 1/4$, $T/m=0.2$,
the formula (rewritten in our conventions)
$1/m\t_{\rm class} = 5.8\, 10^{-6}(6\l T/m^3)^{1.39}$ 
found in this work leads to a relaxation time 
$m\t_{\rm class}\approx 1.3\, 10^5$. 
This is much larger than the $m\t_{cl}\approx 2500$ found here 
in the ``broken phase'' (Sect.\ \ref{sc}).

An interesting question is how the
two time scales 
$\t_{BE}$ and $\t_{cl}$ are related 
to particle scattering and damping. 
A perturbative computation (which includes direct scattering
through the setting-sun diagram), indicates that the damping time 
would be of the order of the BE-relaxation time $\t_{BE}$ (i.e. much shorter
than the relaxation time away from BE behavior). Preliminary numerical results
for the damping time are consistent with these values
%\cite{SaSm00b}.
\cite{SaSm00b_1,SaSm00b_2}.
 
This is a favorable result for the Hartree ensemble method. However the gradual
drift away from a BE distribution and the corresponding cooling of the system
reveals a shortcoming. This is additional to the incorrect prediction by the
Hartree method, of the order of phase transitions. Further improvements are
needed, in particular if large time scales are to be investigated. 

\subsubsection*{Acknowledgments}
We thank Gert Aarts for useful comments.
This research was supported by FOM/NWO.

\bibliographystyle{h-elsevier2}
\bibliography{thermal}

\begin{thebibliography}{10}

\bibitem{MoRu00}
G.D. Moore and K. Rummukainen,
\newblock Phys. Rev. D61 (2000) 105008, hep-ph/9906259,
\newblock %%CITATION = HEP-PH 9906259;%%.

\bibitem{SmTa99}
W.H. Tang and J. Smit,
\newblock Nucl. Phys. B540 (1999) 437, hep-lat/9805001,
\newblock %%CITATION = HEP-LAT 9805001;%%.

\bibitem{afterpreh_1}
S.Y. Khlebnikov and I.I. Tkachev,
\newblock Phys. Rev. Lett. 77 (1996) 219, hep-ph/9603378,
\newblock %%CITATION = HEP-PH 9603378;%%.

\bibitem{afterpreh_2}
T. Prokopec and T.G. Roos,
\newblock Phys. Rev. D55 (1997) 3768, hep-ph/9610400,
\newblock %%CITATION = HEP-PH 9610400;%%.

\bibitem{afterpreh_3}
G. Felder and L. Kofman,
\newblock Phys. Rev. D63 (2001) 103503, hep-ph/0011160,
\newblock %%CITATION = HEP-PH 0011160;%%.

\bibitem{newewbary_1}
J. Garc{\'\i}a-Bellido et~al.,
\newblock Phys. Rev. D60 (1999) 123504, hep-ph/9902449,
\newblock %%CITATION = HEP-PH 9902449;%%.

\bibitem{newewbary_2}
A. Rajantie, P.M. Saffin and E.J. Copeland,
\newblock Phys. Rev. D63 (2001) 123512, hep-ph/0012097,
\newblock %%CITATION = HEP-PH 0012097;%%.

\bibitem{AaBo00}
G. Aarts, G.F. Bonini and C. Wetterich,
\newblock Phys. Rev. D63 (2001) 025012, hep-ph/0007357,
\newblock %%CITATION = HEP-PH 0007357;%%.

\bibitem{AaSm98_1}
G. Aarts and J. Smit,
\newblock Nucl. Phys. B511 (1998) 451, hep-ph/9707342,
\newblock %%CITATION = HEP-PH 9707342;%%.

\bibitem{AaSm98_2}
G. Aarts and J. Smit,
\newblock Phys. Lett. B393 (1997) 395, hep-ph/9610415,
\newblock %%CITATION = HEP-PH 9610415;%%.

\bibitem{CoHa94}
F. Cooper et~al.,
\newblock Phys. Rev. D50 (1994) 2848, hep-ph/9405352,
\newblock %%CITATION = HEP-PH 9405352;%%.

\bibitem{BoVe00}
D. Boyanovsky and H.J. de~Vega,
\newblock (2000), astro-ph/0006446,
\newblock %%CITATION = ASTRO-PH 0006446;%%.

\bibitem{Miea00}
B. Mihaila et~al.,
\newblock Phys. Rev. D62 (2000) 125015, hep-ph/0003105,
\newblock %%CITATION = HEP-PH 0003105;%%.

\bibitem{BeCo00}
J. Berges and J. Cox,
\newblock (2000), hep-ph/0006160,
\newblock %%CITATION = HEP-PH 0006160;%%.

\bibitem{MaWo91_1}
L. Mandel and E. Wolf,
\newblock Optical coherence and quantum optics (Cambridge University Press,
  1995).

\bibitem{MaWo91_2}
J.R. Klauder and B.S. Skagerstam,
\newblock Coherent States (World Scientific, 1985).

\bibitem{SaSm00}
M. Sall{\'e}, J. Smit and J.C. Vink,
\newblock Phys. Rev. D64 (2001) 025016, hep-ph/0012346,
\newblock %%CITATION = HEP-PH 0012346;%%.

\bibitem{SaSm00b_1}
M. Sall{\'e}, J. Smit and J.C. Vink,
\newblock (2000), hep-ph/0008122,
\newblock %%CITATION = HEP-PH 0008122;%%.

\bibitem{SaSm00b_2}
M. Sall{\'e}, J. Smit and J.C. Vink,
\newblock Nucl. Phys. Proc. Suppl. 94 (2001) 427, hep-lat/0010054,
\newblock %%CITATION = HEP-LAT 0010054;%%.

\end{thebibliography}

\end{document}